# Tidal Shear on Density Perturbations and Galaxy Formation

Alejandro González S.

*Instituto Nacional de Astrofísica, Optica y Electrónica, A.P. 51 y 216, C.P. 72000, Tonantzintla, Puebla. México*

**Abstract.**
 The strength of the tidal shear produced by the large-scale density field acting on primordial density perturbations is calculated in power law models. It is shown that the large-scale tidal field could strongly affect the morphology, orientation and angular momentum acquisition of density peaks possibly playing a role in the formation of early-type galaxies. Evidence is presented that the correlation between the orientation of perturbations and the large-scale density field could be a common property of Gaussian density fields with spectral indexes $n < 0$. We argue that alignment of structures as those detected for elliptical galaxies can be used to probe the flatness of the spectrum on large scales but it cannot determine the exact value of the spectral index.

## 1. Correlation of the Local Deformation Tensor of peaks with the Large-scale Density Field

The response of density perturbations to a tidal field depends on their morphology, the strength of the tidal field, and on their relative orientation between the shape and the field. The tidal field exerts a torque on the perturbations, transfering angular momentum to them. The torque per unit mass decreases with the distance as $R^{-3}$, but increases proportionally to the body mass. If the power spectrum of perturbations is flat as suggests the determination on large scale of the spectral index $n = -1.5 \pm 0.5$ by Peacock (1991) and Vogeley et al. (1992), this mass increases as $R^3$ and a divergence in the total torque could appears, i.e. the local properties of the fluctuations can be affected by the large-scale density field. We will study this possibility. The primordial orientation of perturbations is also important because it would determine the angular momentum acquired for the structures in their formation. Perturbations which are born with their major axes nearly aligned with those of the tidal tensor acquire null or small angular momentum. It could be possible that some of these perturbations preserve their initial orientations during the non-linear evolution of the density field. There are some evidences which suggest that the initial number of these peaks could be statistically significant in the past and is preserved today. Studies in this direction have reported anisotropies of the distribution of relative orientations between the major axes of galaxies and that of their host cluster (e.g. Adams, Strom and Strom, 1986) and of the orientations between clusters

1

(West 1989, Plionis et al. 1992, Plionis 1993). Moreover, Lambas, Groth and Peebles (1988, LGP) found not only a trend of the mayor axes of galaxies to be aligned with their host structures (see also Djorgovski, 1986), but also that this effect was morphology dependent: alignments were only detected for elliptical galaxies. LGP suggested that the orientation of ellipticals galaxies could reflect the primordial orientation of maxima of the density field. In such a case, the angular momentum of spiral galaxies need not have any connection with the orientations of ellipticals. Alignment of structures were also detected in the N-body simulations by Barnes and Efstathiou (1987) who pointed out, "*the most striking effect seen in these tests is the tendency for nearby objects to point at each other. It seems likely that objects are born with these orientations; if they had been formed with random major axes and sheared into line by tidal torques, we would expect a strong spin-vector versus separation-vector effect, which we do not find*". Could those perturbations aligned with the tidal field be the progenitors of elliptical galaxies ?

### 1.1. Strength of the tidal field

Once the density field is generated within a $64^3$ cubic grid, and smoothed with a Gaussian filter function, the peaks mass grows as (Bardeen et. al. 1986)

$$M(R_f) = (2\pi)^{3/2} \bar{\rho} R_f^3 = 4.37 \times 10^{12} R_f^3 h^{-1} M_\odot. \tag{1}$$

For a galactic mass scale $M \approx 10^{12} M_\odot$, $R_f \approx 0.6 \text{Mpc} \equiv R_g$. Suppose that the density field is newly smoothed on a larger scale by using a filter radius $R_f^{(2)}$. If a change in the elements of the deformation tensor (Zel'dovich, 1970)

$$\mathbf{D}_{jk} = \delta_j k + b(t) \frac{\partial v_k}{\partial q_j}, \tag{2}$$

is observed, then it can be attributed to the ramaining superposition of density waves bounded by the spherical shell defined by the radii difference $\Delta R_f = R_f^{(2)} - R_g$. Intuitively it is expected that when $R_g \approx R_f^{(2)}$, the general properties of a given density peak will change only slightly, i.e. there exist a correlation on those scales, between the peak with the surrounding distribution of mass. Let us now resort to a widely used concept in theory of matrix perturbation (Stewart and Sun 1990). This theory bounds the changes in physical processes described by a matrix when the elements of the matrix change. The theory assess, e.g. how far the "magnitude" of the deformation tensor $\mathbf{D}_{ij}^{(1)}$ will change when its elements change as $\mathbf{D}_{ij}^{(2)} = \mathbf{D}_{ij}^{(1)} + \mathbf{\Delta}_{ij}$, due to a perturbation, $\mathbf{\Delta}_{ij}$. A prerequisite for answering this question is to make precise the term "magnitude", which is done trough of defining a norm in the matrixes space. We have adopted the Frobenius norm

$$\mathbf{F} \equiv \parallel \mathbf{\Delta}_{ij} \parallel \equiv \sqrt{\sum_{i,j} \mid \mathbf{D}_{ij}^{(2)} - \mathbf{D}_{ij}^{(1)} \mid^2}. \tag{3}$$

When $\mathbf{D}_{jk}$ does not suffer important changes $\mathbf{F}$ will tend to take small values. The lower limit correspond to $\mathbf{F} = 0$ in which the elements of the deformation tensor are not affected.

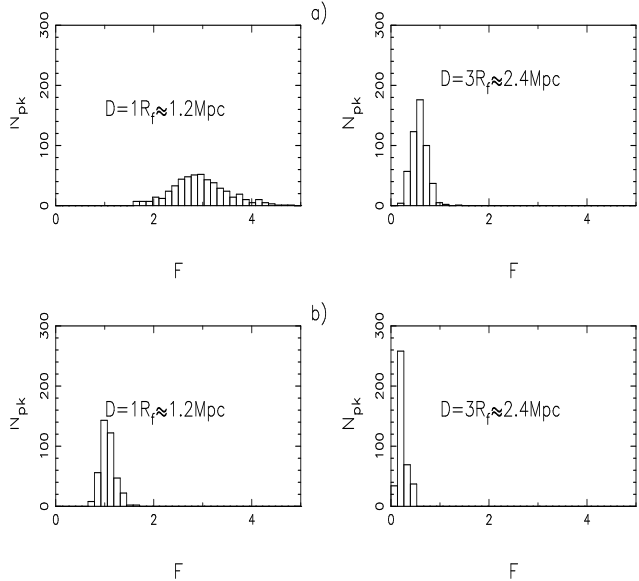

Figure 1. The Frobenius norm as a function of the distance to the peaks center. a) Top panels correspond with a spectral index $n = -2$, whereas b) bottom panels correspond with $n = -1$.

## 2. Results

We will be considering an Einstein-de Sitter ($\Omega = 1$) Universe. From Figure 1, which displays the distribution of peaks with norm F, a clear difference in the strength of the tidal shear is observed between the different models. The fact that the maximum of the distribution of peaks in the cases $n = -1$ and $n = -2$ shows a considerably deviation from zero at large radii of filtering, confirms the influence of the superposition of the density waves which conform the large-scale perturbations on the local deformation of small-scale perturbations. This is seen from Figure 2, where we compare the strength of the influence on the density peaks as a function of the filtering scale. The norm of the deformation tensor at $R_f = 1$ was used to normalize the maximum effect to unity. Steep spectra, $n = 1, 0$, show a correlation of the peaks only with the nearest distribution of mass on scales smaller than $R_f = 2$. Flat spectra show important effects coming from even larger-scales, $R_f \approx 4-5$. An insight of the decaying rate of the effect on density peaks is provided by the difference of F's between different shells which we have bined as $\Delta F = F(R_f = 2) - F(R_f = 1), F(R_f = 4) - F(R_f = 2), F(R_f = 8) - F(R_f = 4)$. It is observed from Figure 2, a general tendency for the influence to decay. However, the rate of dacaying is dramatically faster for the model with steep spectra than for those with flatter ones.

These results suggest that tidal shear induced by large-scale inhomogeneities can significantly affect the morphology and orientation of those primordial density peaks whose inertia and deformation tensors are misaligned. The situation will be different for the large per cent of the peaks for which both tensor are aligned. They can be considered as candidates to present alignment with the larger-scale perturbations.

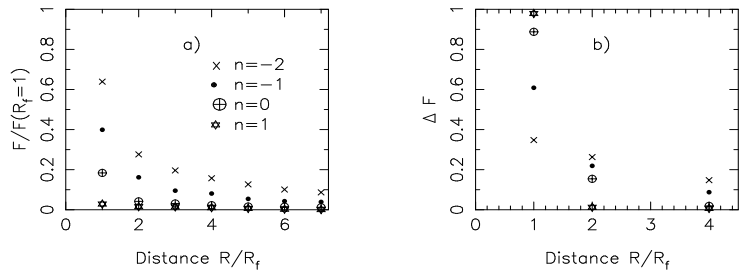

Figure 2.  a) An estimate of the strength of the tidal field as a function of the mass scale of filtering. b) The rate of decaying of the contribution to the tidal effects of the mass shells.

Concerning primordial alignments, we found (González, 1995a,b) that for flat spectra there exist approximatelly 35% of the peaks whose major axes of the inertia tensor is parallel to the main axes of the density peaks of cluster mass, $M \approx 10^{15} M_\odot$ with $R_g \approx 10 Mpc$. In other words, information of the morphology of large perturbations is apparently preserved in their inner regions, in a similar way to the observed alignment of cD galaxies with the main axes of the host cluster (West, 1994). We shall further study whether alignment of these structures survive to the non-linear evolution of the density field.

## References

Adams, M.T., Strom, K.M. & Strom, S.E., 1980, ApJ, 238, 445.
Barnes, J.A. & Efstathiou, G., 1987, ApJ, 319,575.
Djorgovski, S., 1986. In *Nearly Normal Galaxies*, Ed. S. Faber. New York Springer-Verlag, 226.
Bardeen, J.M., Kaiser, N., Bond, R. & Szalay, R., 1986, ApJ, 304, 15.
González A., 1994, Ph.D. Thesis. University of Sussex, UK.
González A., 1995a, Submitted to A&A
González A., 1995b, Submitted to ApJ
Lambas, D.G., Groth, E.J. & Peebles, P.J.E., 1988. AJ, 95, 996.
Peacock, J.A., 1991, MNRAS, 253, 1P.
Plionis, M., Valdarnani, R. & Jing, Y., 1992, ApJ, 398, 12.
Plionis, M., 1993, ApJ, preprint.
Stewart, G.W. & Sun, J., 1990. Matrix Perturbation Theory. Academic Press.
Vogeley, M.S., Park, C., Geller, M.J., Huchra, J.P., 1992, ApJ, 391, L5.
West, M.J., 1989, ApJ, 347, 610.
West, M.J., 1994, MNRAS, 268, 79.
Zel'dovich, Ya. B., 1970, A&A, 5, 84.